\documentclass[preprint,aps,amsmath,amssymb,showpacs,pre]{revtex4}
\usepackage{graphicx}
\begin{document}
\title{Solution of a model of SAW's with multiple monomers per site
  on the Husimi lattice} 
\author{Tiago J. Oliveira}
\email{tiagojo@if.uff.br}
\author{J\"urgen F. Stilck}
\email{jstilck@if.uff.br}
\affiliation{Instituto de F\'{\i}sica\\
Universidade Federal Fluminense\\
Av. Litor\^anea s/n\\
24210-346 - Niter\'oi, RJ\\
Brazil}
\author{Pablo Serra}
\email{serra@famaf.unc.edu.ar}
\affiliation{Facultad de Matem\'atica, Astronom\'{\i}a y F\'{\i}sica\\
Universidad Nacional de C\'ordoba\\
C\'ordoba - RA5000\\
Argentina}
\date{\today}

\begin{abstract}
We solve a model of self-avoiding walks which allows for a site to be
visited up to two times by the walk on the Husimi lattice. This model
is inspired in the Domb-Joyce model and was proposed to describe the
collapse transition of polymers with one-site interactions only. We
consider the version in which immediate self-reversals of the walk are
forbidden (RF model). The phase diagram we obtain for the
grand-canonical version of the model is similar to the one found in
the solution of the Bethe lattice, with two distinct polymerized
phases, a tricritical point and a critical endpoint. 
\end{abstract}

\pacs{05.40.Fb,05.70.Fh,61.41.+e}

\maketitle

\section{Introduction}
\label{intro}
Although linear polymers in solution may be studied using continuous
models, much has been learned about these systems through models of
self- and mutually avoiding walks placed on lattices \cite{f66}. The
excluded volume constraint makes it quite difficult to treat these
problems, even trying to answer apparently simple questions such as
the number of walks 
with a given number of steps, which is relevant in series expansion
approaches to the problem, is a challenging and rich field of research
(for a recent work in this field see \cite{c05}). The connection of
such models with the $n$-vector model of magnetism in the formal limit
$n \to 0$ \cite{dg72} and the application of the ideas of scaling and
the renormalization group to polymer systems \cite{dg79} have also
been central in the development of this area of research.

If the polymer chain is placed in a poor solvent, as the temperature
is decreased eventually the chain changes from an extended
configuration (in which the entropy is favored) to a collapsed
configuration (with less contact between the polymer and the
solvent and thus a smaller energy). The temperature where this
transition happens was called $\theta$ temperature \cite{f66}. Although
this transition may be modeled using lattice models where the solvent
is included explicitly \cite{w84}, these models in a certain limit
lead to a simpler model which has become the standard model for this
phenomenon, in which, besides the repulsive excluded volume
interactions, attractive interactions between monomers on first
neighbor sites but not consecutive along the chain are introduced
(ISAW model). The configurations in this model are self-avoiding walks 
whose steps link monomers located on first neighbor lattice sites.
The collapse transition in  models which 
are grand-canonical with respect to the number of monomers in the
system usually appears as a tricritical point in the phase
diagram. In the fugacity vs. temperature phase diagram, a
non-polymerized phase is present in the region of low monomer
fugacity, and a polymerized phase is stable for higher 
fugacities. The transition between these phases is of first order at
low temperatures and becomes continuous at higher temperatures. These
regimes are separated by a tricritical point, where the collapse
transition occurs. The ISAW model has been extensively studied on the square
lattice \cite{ds85}, and the exact tricritical exponents of the
diluted polymer model were found \cite{ds87}. Under certain
conditions, an even richer phase diagram is found, with the presence
of a dense polymerized phase at finite monomer fugacity, where the
density of empty lattice sites vanishes. This additional phase was
found in solutions of the model on $q=4$ Husimi lattices with
interactions between {\em bonds} \cite{sms96,p02a} of the polymer located
on opposite sides of elementary squares of the lattice, as well as in cluster
approximations of similar models on the square lattice
\cite{bp00}. Transfer matrix calculations on strips of finite widths
support the existence of this phase on the square lattice as well
\cite{mos01}. The dense phase is absent if the interactions are only 
between monomers, but even in this case there are indications that 
this phase will be present if the polymer chains are sufficiently
stiff \cite{p02}.

Usually, the introduction of interactions in the polymer models is
a source of difficulty both in approximate solutions and in 
transfer-matrix approaches. Thus, a model introduced recently for
studying the collapse transition in polymers where only {\em one-site}
interactions are present is quite interesting \cite{kpor06}. This 
model allows for multiple occupancy of a site by up to $K$ monomers,
assuming that the attractive interactions are restricted to monomers 
which occupy the same site of the lattice. One way to justify the 
model would be to discretize the original system of a polymer in a 
solution  using a regular lattice, and choosing the size of the elementary
cell of this lattice to be big enough to accommodate up to $K$ monomers 
of the polymer. Also, the length of the bonds has to be larger than the 
size of the cells, so that two monomers in the same cell will never be
connected by a bond. Two versions of the model were studied by extensive
numerical simulations in \cite{kpor06} on the square and on the cubic lattice.
In the RA (immediate reversals allowed) model, there are no restrictions on 
the walks on the lattice, while in the RF model (immediate reversals
forbidden)  
only a subset of the possible walks is considered: those in which the walk 
does not return to the original site immediately after reaching a new site. In 
the simulations done for the RF model on the cubic lattice for $K=3$,
a transition 
between extended and collapsed polymerized phases is found. 

Recently, the RA and RF 
models were solved on a Bethe lattice for $K=2$, in order to compare their 
thermodynamic behaviors with the much studied ISAW model regarding the usual
collapse transition \cite{ss07}. The solution of
the RF model 
on the Bethe lattice produces a phase diagram in which the polymerization 
transition remains continuous for small non-zero values of $\omega_2$, becoming
of first order at higher values of this statistical weight, a behavior
which is  
also found in the ISAW model. Besides the regular polymerized phase, a
second phase 
is stable for sufficiently high values of $\omega_2$ and low values of
$\omega_1$ 
where only empty and double-occupied sites are present. 

Here the RF model is solved on a $q=2(\sigma+1)$ Husimi lattice, which is the
central region of  
a Cayley tree built with squares. There are $\sigma+1$ squares
incident on each lattice site. The thermodynamic behavior of models on such 
a lattice is expected to be closer to the one obtained on a regular
lattice with 
the same coordination number, and we were motivated for this calculation mainly
for two reasons. Although the RF model on the Bethe lattice displays a
tricritical 
point which may be associated to the usual collapse transition of
polymers, 
if we parametrize the model such that $\omega_1=z$, the activity of a monomer,
and $\omega_2=z^2\omega$, where $\omega$ is the Boltzmann factor associated to
a pair of interacting monomers, the tricritical point is found at $\omega<1$, 
a value that corresponds to a {\em repulsive} interaction between monomers at
the same site. We are interested in finding out if a calculation which should
lead to results closer to the ones on regular lattices might shift the 
tricritical point to the region in the parameter space which corresponds to 
attractive interactions. Also, it is of interest to find out if the second 
polymerized phase is still present in the phase diagram of the model 
on the Husimi lattice. We found that actually with a convenient parametrization
the tricritical point is located in the physically expected region and that
the second polymerized phase is still present in the Husimi lattice solution,
although it occupies a smaller region of the parameter space than the one
found for the Bethe lattice.

In section \ref{ms} we define the model in more detail and present its
solution on the Husimi lattice. Final discussions and the conclusion may
be found in section \ref{con}.

\section{Definition of the model and solution on the Husimi lattice}
\label{ms}
We consider a Husimi tree, a Cayley tree built with polygons, which
in our case will be squares. Sometimes this tree is also called a cactus. 
As also happens for the Cayley tree, in the thermodynamic limit the
fraction of sites which are on the surface of the tree does not vanish, and
this accounts for the fact that the solution of models on such trees usually
shows a behavior which does not resemble the one found on regular lattices. If,
however, the behavior of models in the central region of the trees is
considered,  
for many models the exact solution corresponds to the Bethe 
approximation of the same model on a regular lattice with the same
coordination  
number, and this is the reason why this is called a Bethe lattice solution 
\cite{b82}. The Husimi lattice corresponds to the central region of a Cayley
tree built with polygons (squares in our case), and since closed paths are
present (although  
restricted to single elementary squares), it is expected that the solution
of models on this tree will be closer to the one found on regular
lattices, and this is confirmed in many cases.

\begin{figure}
\begin{center}
\includegraphics[height=6.0cm]{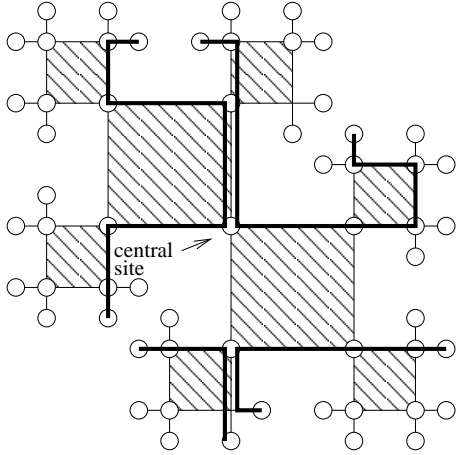}
\caption{Example of a configuration for the $K=2$ RF model on a
  Husimi tree with square ramification $\sigma=1$ and
  2 generations.  Polymer bonds (steps of the walks) are
  represented by thick lines, while the lattice bonds are thin
  lines. The weight of this configuration is $\omega_1^{12}\omega_2^4$.} 
\label{f1}
\end{center}
\end{figure}

\begin{figure}[h]
\begin{center}
\includegraphics[height=6.0cm]{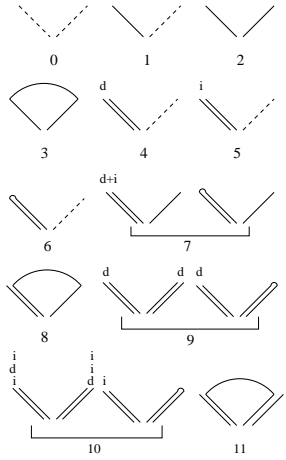}
\caption{Root configurations of the RF model with $K=2$ on the Husimi
  lattice.}  
\label{f2}
\end{center}
\end{figure}

The allowed configurations of the $K=2$ RF model are walks that may
visit a site  
one or two times, with their initial and final monomers placed only on
the surface 
of the tree. Also, as stated above, immediate reversals of the walk
are forbidden. 
In figure \ref{f1} a possible configuration is shown on a tree with two
generations of squares. The statistical weight of a configuration will be 
$\omega_1^{N_1}\omega_2^{N_2}$, where $N_1$ and $N_2$ are the numbers of sites 
with one and two monomers, respectively. For simplicity, the surface
of the tree was chosen to be defined by sites connected to a single
site of the first generation of squares. To solve the model on the
Husimi tree, we consider rooted subtrees, and define partial partition
functions for these trees, for fixed configurations of the bonds
incident on the root
site. The operation of attaching three sets of $\sigma$
$n$-generations subtrees to a new root square will result in a
$n+1$-generations subtree, leading to recursion relations for the
partial partition functions. 
Notice that we decided to consider the monomers
placed on the same site to be {\em indistinguishable}, in opposition to
what was adopted in the solution on the Bethe lattice \cite{ss07},
where they were supposed to be distinguishable. This new convention
was motivated mainly by two aspects: as already observed in the discussion
of the Bethe lattice solution, the tricritical point of the $K=2$ RF
model is located in the region in the parameter space which
corresponds to {\em repulsive} interactions between monomers. If two
monomers placed on 
the same site are considered to be indistinguishable, the weight
$\omega_2$ will be multiplyed by a factor of 2, and this will
shift the tricritical point towards the region of attractive
monomers. Also, in the original simulations \cite{kpor06} the
configurations of a 
walk on the lattice are labeled by the sequence of sites visited by
the walk, and this convention corresponds to indistinguishable
monomers. Although we do no detailed comparisons of our results with
the simulations, since they were done for $K=3$, we decided to adopt
the same convention.
In figure \ref{f2} the 11 configurations of
the bonds incident on the root site of subtrees are depicted. For
brevity, only one configuration of pairs related by reflection
symmetry is shown. For example, there are a total of 4 configurations
with the same partial partition function in group 7 of the figure. We
notice that the connection 
of the incoming bonds to the monomers placed on the root site is not
fixed. Since we assume the monomers to be indistinguishable, to
write down the recursion relations for the partial partition functions
we must separate incoming double bonds in two groups: if both bonds
visited the same sites since the boundary of the tree they are labeled
as {\em i} (as in configuration 5), otherwise, they are labeled as
{\em d} (as in configuration 4). This information is essential to
allow us to correctly determine the  multiplicity of the
contributions to the recursion relations below.

\begin{figure}[h]
\begin{center}
\includegraphics[height=6cm]{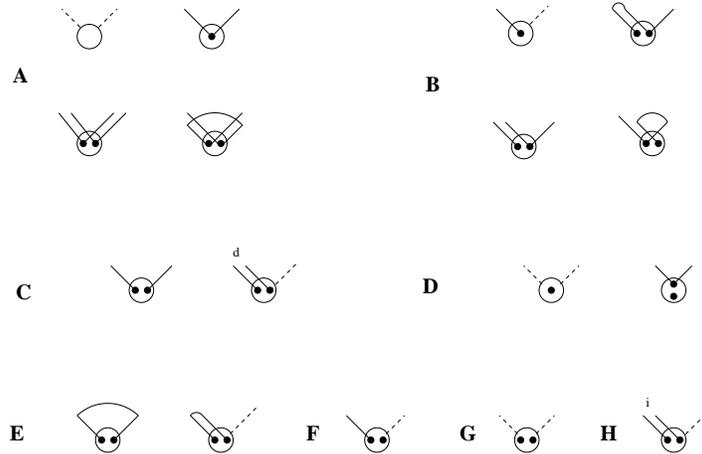}
\caption{Contributions to the vertex functions. The monomers are
  represented by dots.}   
\label{f3}
\end{center}
\end{figure}

We may now proceed obtaining the recursion relations for the partial
partition functions.  
We notice that 
certain partial partition functions appear in the recursion relations
only in linear 
combinations. Thus, $g_3$ and $g_6$ always appear as $g_3+g_6$, and the
other linear combinations are $2g_7+g_8$ and $2g_9+g_{11}$. We therefore
may reduce the number of independent variables in the recursion relations by
3. The partial partition functions usually diverge in the thermodynamic
limit, thus we consider the 8 ratios defined below, which often remain
finite, and in the
thermodynamic limit a phase of the system may be 
associated to a fixed point of the recursion relations for the ratios.

\begin{subequations}
\begin{eqnarray}
R_1 &=& \frac{g_1}{g_0},\\
R_2 &=& \frac{g_2}{g_0},\\
R_3 &=& \frac{g_3+g_6}{g_0}, \\
R_4 &=& \frac{g_4}{g_0}, \\
R_5 &=& \frac{2g_7+g_8}{g_0}, \\
R_6 &=& \frac{2g_9+g_{11}}{g_0}, \\
R_7 &=& \frac{g_5}{g_0}, \\
R_8 &=& \frac{g_{10}}{g_0}.
\end{eqnarray}
\end{subequations}
It is now
convenient to define first some linear combinations of partial
partition functions which appear repeatedly in the recursion
relations, whose contributions
are shown graphically in figure \ref{f3} for the particular
ramification of squares $\sigma=1$. They are:
\begin{subequations}
\begin{eqnarray}
A &=& g_{0}^{\sigma} \left[ 1 + \sigma \omega_1 R_2
  +\binom{\sigma}{2}\omega_{1} R_{1}^2 
  +\sigma\omega_2R_6+\sigma\omega_2R_8
  +2\binom{\sigma}{2}\omega_2R_4^2
  +2\binom{\sigma}{2}\omega_2R_4 R_7\right. \nonumber \\
&&+\binom{\sigma}{2}\omega_2R_7^2
  +4\binom{\sigma}{2}\omega_2R_3R_4
  +2\binom{\sigma}{2}\omega_2R_3R_7
  +4\binom{\sigma}{2}\omega_2R_2R_4
  +2\binom{\sigma}{2}\omega_2R_2R_7\nonumber \\ 
&&+4\binom{\sigma}{2}\omega_2R_2R_3
  +3\binom{\sigma}{2}\omega_2R_2^2
  +2\binom{\sigma}{2}\omega_2R_1R_5
  +9\binom{\sigma}{3}\omega_2R_1^2 R_2\nonumber \\
&&\left. +6\binom{\sigma}{3}\omega_2R_1^2R_3
  +6\binom{\sigma}{3}\omega_2R_1^2R_4
  +3\binom{\sigma}{3}\omega_2R_1^2R_7
  +3\binom{\sigma}{4}\omega_2R_1^4 \right]\\
B &=& g_{0}^{\sigma} \left[\sigma\omega_1R_1
  +6\binom{\sigma}{2}\omega_2R_1R_2
  +4\binom{\sigma}{2}\omega_2R_1R_3
  +4\binom{\sigma}{2}\omega_2R_1R_4
  +2\binom{\sigma}{2}\omega_2R_1R_7  \right. \nonumber \\
&& \left. +3\binom{\sigma}{3}\omega_2R_1^3
  +\sigma\omega_2R_5 \right] \\
C &=& g_{0}^{\sigma} \left[\binom{\sigma}{2}\omega_2R_1^2
  +\sigma \omega_2R_2
  +\sigma\omega_2R_4 \right] \\
D &=& g_{0}^{\sigma} \left[ \omega_1+\binom{\sigma}{2}\omega_2R_1^2
  +\sigma\omega_2R_2 \right] \\
E &=& g_{0}^{\sigma} \left[ \sigma \omega_{2} R_3 \right] \\
F &=& g_{0}^{\sigma} \left[ \sigma \omega_{2} R_{1} \right] \\
G &=& g_{0}^{\sigma} \left[ \omega_{2} \right] \\
H &=& g_{0}^{\sigma} \left[\sigma \omega_2 R_7 \right]
\end{eqnarray}
\end{subequations}

\begin{figure}[h]
\begin{center}
\includegraphics[height=6cm]{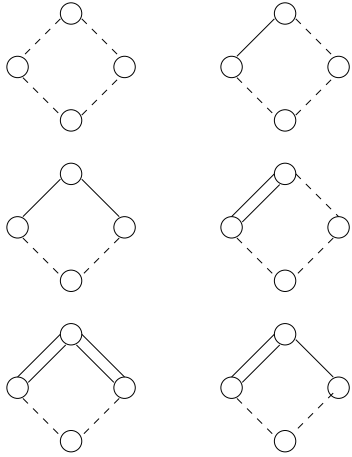}
\caption{Contributions to the recursion relation for the partial
  partition function $g_0$..}   
\label{f4}
\end{center}
\end{figure}

Now we may proceed considering the operation of attaching three
sets of $\sigma$ subtrees to a new root square, summing all possible
contributions for 
a fixed configuration of the bonds which are incident on the new root site. In
figure \ref{f4} the 
graphical representation of the contributions to $g^\prime_0$ are
shown to illustrate this process, in the order they appear in the
equation below. The recursion relations are:
 \begin{subequations}
\begin{eqnarray}
g_0^\prime&=&A^3+2AB^2+B^2(2C+H+D+2E)+(2A+G)(H^2+2HC+2C^2+4CE+2EH)\nonumber \\
&& +2BF(2C+H+2E); \\
g_1^\prime &=& 2A^2B+2B^3+2AB(2C+H+D+2E)+2B(2C+H+D+2E)^2 \nonumber \\
&&+2AF(2C+H+2E)+2FG(2C+H+2E)+2B(H^2+2HC+2C^2+4CE+2EH) \nonumber \\
&&+2F(2C+H+D+2E)(2C+H+2E)+4BF^2;\\
g_2^\prime &=&
AB^2+2B^2(2C+H+D+2E)+(2C+H)^3+3(2C+H)^2(D+2E) \nonumber \\
&&+3(2C+H)(D+2E)^2+2BF(2C+H+2E)+F^2G+4F^2(2C+H) \nonumber \\
&&+2F^2(D+2E);
\label{g2p}\\
g_3^\prime
&=&(D+2E)^3+2F^2(D+2E)+F^2G;\\ 
g_4^\prime &=& 2A^2C+2ABF+2B^2C+2BF(2C+H+D+2E)+2ACG+2CG^2+\nonumber \\
&& 2F^2(2C+H+2E)+2BFG+2C(H^2+2HC+2C^2+4CE+2EH); \\ 
g_5^\prime &=& 2H(A^2+B^2+AG+G^2+H^2+2HC+2C^2+4CE+2EH); \\
g_6^\prime &=& 2A^2E+2B^2E+2AEG+2EG^2+ \nonumber \\
&& 2E(H^2+2HC+2C^2+4CE+2EH); \\
g_7^\prime &=& 2AB(C+H+E)+2B^2F+2B(C+H+E)(2C+H+D+2E)+2BG(C+H+E)+
\nonumber \\
&&2F(C+H+E)(2C+H+2E)+2F[(2C+H)^2+2D(2C+H)+4E(2C+H)] \nonumber \\
&&+2FG(2C+H)+2F^3; \\
g_8^\prime &=& 2F(D+2E)^2+2FG(D+2E)+2FG^2+2F^3; \\
g_9^\prime &=& AC(C+2E)+2BF(C+E)+2CG(C+2E)+F^2(2C+H); \\
g_{10}^\prime &=& AH(2C+H+2E)+2BFH+2GH(2C+H+2E); \\
g_{11}^\prime &=& F^2(D+2E+2G).
\end{eqnarray}
\end{subequations}
In these recursion relations the need to distinguish between the cases $i$ and
$d$ of double bonds in the definition of the partial partition functions is
apparent, for example, in the contributions to $g_2^\prime$: the terms
proportional to $C$ (case $d$) have an additional factor 2 as compared to the
terms proportional to $H$ (case $i$).

Finally, we will consider the operation of attaching $\sigma+1$
subtrees to 
the central site of the lattice. This operation is similar to the ones
realized to obtain the recursion relations and the result is:
\begin{eqnarray}
Y &=& g_{0}^{\sigma+1} \left[ 1 + \binom {\sigma+1}{2} \omega_{1}
  R_{1}^2 + 3 \binom 
{\sigma+1}{4} \omega_{2} R_{1}^4 + 9 \binom {\sigma+1}{3} \omega_{2}
  R_{1}^2 R_{2} + 6
\binom {\sigma+1}{3} \omega_{2} R_{1}^2 R_{3} \right. \nonumber \\ 
&&\left. + 3 \binom {\sigma+1}{3} \omega_{2} R_{1}^2 (2R_{4}+R_7) + 2 \binom 
{\sigma+1}{2} \omega_{2} R_{1} R_{5} \right. + (\sigma+1) \omega_{1}
  R_{2} + 3 \binom {\sigma+1}{2} 
  \omega_{2} R_{2}^2\nonumber \\
&&\left.  + 4 
\binom {\sigma+1}{2} \omega_{2} R_{2} R_{3} + 2 \binom {\sigma+1}{2}
  \omega_{2} R_{2} (2R_{4}+R_7)  
+ 2 \binom {\sigma+1}{2} \omega_{2} R_{3} (2R_{4}+R_7)  \right. \nonumber \\
&& \left. + \binom
  {\sigma+1}{2} 
\omega_{2} (2R_{4}^2+R_7^2)+2\binom {\sigma+1}{2} \omega_{2} R_{4}R_7
+ (\sigma+1) \omega_{2} (R_{6}+R_8) \right].
\end{eqnarray}
It is now easy to obtain the probabilities of single and double
occupancy of the central site. The results are
\begin{eqnarray}
\rho_1&=&\omega_1\frac{\binom {\sigma+1}{2}
  R_{1}^2+(\sigma+1)R_{2}}{D},\\
\rho_2&=&1-\rho_1-\frac{1}{D},
\end{eqnarray}
where $D=Y/g_0^{\sigma+1}$.

\begin{figure}[h!]
\begin{center}
\includegraphics[height=6.0cm]{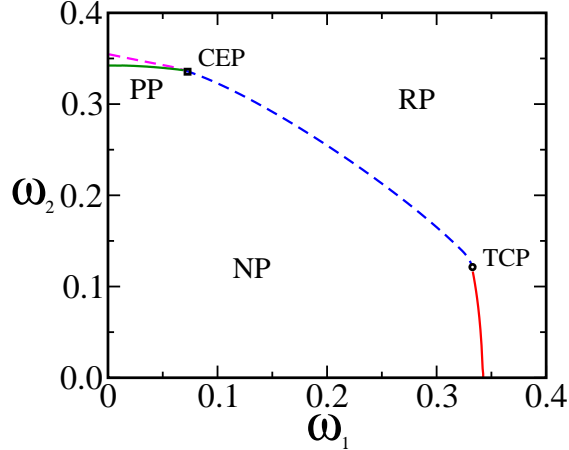}
\caption{(color on line) Phase diagram of the model on the Husimi
  lattice. Full lines 
  are continuous transitions and dashed lines are first order
  transitions between the NP and RP (blue) phases and between the PP and RP
  (purple) phases. The tricritical point and the critical enpoint are also shown.}  
\label{f6}
\end{center}
\end{figure}

As stated above, to study the thermodynamic behavior of the model we
have to find the fixed point of the recursion relations, given the
statistical weights $\omega_1$ and $\omega_2$. As expected, the ratios
$R_7$ and $R_8$ vanish for all values of the statistical weights.
On the Husimi lattice a double chain which starts on the
boundary may end at any step, by entering into a square and circulating it
(this corresponds to the contribution $3HD^2$ in the recursion relation
\ref{g2p} for $g_2$). At low values of the 
statistical weights, a non-polymerized phase, characterized by
$\rho_1=\rho_2=0$, is stable. 
At the fixed point associated to this phase all ratios vanish, with
the exception of $R_3$. From the recursion relations above
we may find the following  equation for this fixed point:	 
\begin{equation}
(\omega_1 + 2\sigma\omega_2 R_3)^3 +[2\sigma\omega_2(1 +\omega_2+\omega_2^2)
-1]R_3 = 0.
\end{equation}
At the
stability limit of the NP phase the largest eigenvalue of the jacobian
of the recursion relations is equal to one, and this condition allowed
us to obtain the region of the parameter space where the NP phase is
stable. For large values of the statistical weights, a regular
polymerized (RP) phase is stable, in which the ratios $R_1,R_2,\ldots,R_6$ are
non-vanishing at 
the fixed point. Finally, for small values of $\omega_1$, between the
regions where the NP and RP phases are stable, a third phase appears,
for which all ratios which correspond to an odd number of incoming
bonds at the root site $(R_1,R_5)$ vanish.  We
will call this phase PP (pair 
polymerized). In figure \ref{f6} the phase diagram of the model
is presented for a lattice with $q=4$, and we notice that the transition
between the NP and RP 
phases may be of first or second order, with a tricritical point
located at $\omega_1=0.3325510(6)$ and $\omega_2=0.120544(4)$. The
region where the PP phase is the most stable one is rather small and
the transition between the NP and PP phases is continuous. This
transition line ends at a critical endpoint, which is located at
$\omega_1=0.0695605(5)$ and $\omega_2=0.3370740(2)$. A discontinuous
transition separates the two polymerized phases. The first order
lines were obtained directly from the recursion relations, starting
the iterations with `natural' initial conditions \cite{p02a}. 
In the present calculations, we considered the surface of the tree to
be formed by sites connected to a single site of the next generations,
as shown in figure \ref{f1}. This choice leads to the following
initial values: 
\begin{subequations}
\begin{eqnarray}
R_1&=&2\omega_1;\\
R_2&=&\omega_1^2;\\
R_3&=&0;\\
R_4&=&0;\\
R_5&=&4\omega_1\omega_2;\\
R_6&=&0;\\
R_7&=&2\omega_2;\\
R_8&=&\omega_2^2.
\end{eqnarray}
\end{subequations}
As mentioned above, the ratios $R_7$ and $R_8$ vanish in the termodynamic
limit, but have nonzero values for finite lattices. If we change slightly
the initial conditions, assuming the monomers placed on the surface of the
tree to be distinguishable, these
ratios will vanish identically and the initial value for $R_4$ would be equal
to $2\omega_2$, with no change in the thermodynamic properties
of the model within our numerical precision.

A model which is very similar to the one we are studying here was investigated
recently by R. A. Zara and M. Pretti \cite{zp07} to study the properties of
RNA-like molecules, and actually the phase diagram for the Husimi lattice
solution of this model is similar to the one we obtain here.
It may also be mentioned that no dense
phase, as the ones found in some versions if the ISAW model
\cite{sms96,p02a}, is stable in finite regions of the parameter
space. Actually, such a phase is stable in a region of the
$\omega_1=0$ line, but this fixed point is never reached if $\omega_1
\neq 0$ and therefore is of no physical relevance. 

\section{Final discussions and conclusion}
\label{con}
\begin{figure}[h!]
\begin{center}
\includegraphics[height=6.0cm]{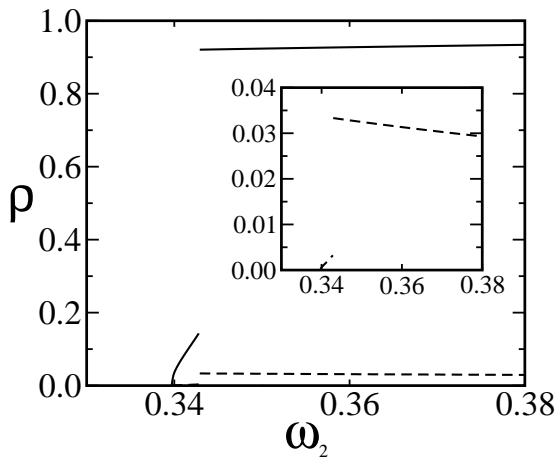}
\caption{Density of sites occupied by one ($\rho_1$, dashed lines) and
two ($\rho_2$-full lines) monomers, as functions of $\omega_2$ for
$\omega_1=0.05$.}  
\label{f7}
\end{center}
\end{figure}

Qualitatively, the phase diagram presented here is similar to the one
found for the RF model on the Bethe lattice
(figure 3 of reference \cite{ss07}). We notice that on the Bethe lattice the second
order line between the NP and the RP phases is located at
$\omega_1=1/3$, while the NP-PP transition happens at
$\omega_2=1/6$ ($\omega_2=1/3$ if the monomers are considered to be
indistinguishable). Those lines are no longer parallel to one of the axes 
for the Husimi lattice solution. The location of the TCP is not
changed much in the two solutions, although the value of $\omega_2$ for
this point on the Husimi lattice solution is about 10 \% larger. The
CEP, although localized at 
almost the same value of $\omega_2$ in the two solutions, shows a much
smaller value of $\omega_1$ on the Husimi lattice. As a consequence,
the area in which the PP phase is stable in the parameter space
is much smaller on the Husimi lattice solution, it is an open question
if this unusual phase appears if the model is considered on regular
lattices. On the Bethe lattice solution the PP phase was called double
occupancy polymerized phase, since $\rho_1$ was found to vanish in
this phase. We notice that in the Husimi lattice solution the density
of sites occupies by a single monomer does not vanish in this phase,
as may be appreciated in the inset of the figure \ref{f7}, where both
densities are 
shown as functions of $\omega_2$ for a fixed value of $\omega_1$.

The comparison of the model with multiple monomers per site (MMS) with the
usual ISAW model is not straightforward. When $\omega_2=0$, the MMS
model corresponds to the ISAW model without attractive
interactions. However, the MMS model for $K=2$ without attractive
interactions corresponds to the line $\omega_2=\omega_1^2$, since we
should associate a statistical weight equal to the activity $z$ to
each monomer placed on the lattice. In the ISAW model a subset of the
walks considered in the MMS model is allowed.  If we actually imagine
this model to be an 
effective description of a continuous model treated in a cell
approximation, we might think the parameter $\omega$ to be the
effective interaction found integrating the position of the two
monomers inside the cell they occupy. If this interpretation is
adopted, the statistical weight of a site occupied by one monomer
would be $\omega_1=z$, where $z$ is the activity of a monomer, while a
double occupied site would contribute with a factor $\omega z^2$ to
the grand canonical partition function, where $\omega$ is the Boltzmann
factor associated to the interaction of two monomers. With this
parametrization, the tricritical point will be located at
$\omega\approx1.09$, which corresponds to attractive interactions between
monomers at the same site. In
the Bethe lattice solution for indistinguishable monomers, the
tricritical point is located at 
$\omega=1$, which corresponds to no interaction.

We also did the calculations of the model for distinguishable monomers, as was
done initially for the Bethe lattice. Within our numerical approximation, this
solution leads to a phase diagram which differs from the one presented here by
a factor of 2 in the values of $\omega_2$. This may be understood by noting
that the recursion relations for the model with distinguishable monomers are the
same ones presented here except for this factor 2 in each term proportional
to $\omega_2$ for $R_1,R_2,\ldots,R_6$ (configurations 5 and 10 are absent in
this case), and that the additional ratios $R_7$ and $R_8$ vanish in the fixed
point for the model with indistinguishable monomers, as stated above.

\section*{Acknowledgements}

This work was partially financed by the Brazilian agency CNPq. PS
acknowledges partial financial 
support  of the Argentinian agencies SECYTUNC and CONICET, and thanks
the Universidade Federal Fluminense for the hospitality.

\end{document}